\documentclass[prl,twocolumn,showpacs,superscriptaddress,nofootinbib]{revtex4}
\usepackage{epsf}
\usepackage{epsfig}
\usepackage{amsmath}

\usepackage{epstopdf}

\usepackage{mathpazo}   

\def\beq{\begin{equation}}
\def\eeq{\end{equation}}
\def\bea{\begin{eqnarray}}
\def\eea{\end{eqnarray}}
\def\beqa{\begin{equation}\begin{array}{l}}
\def\eeqa{\end{array}\end{equation}}

\begin{document}

\title{Empirical transverse charge densities in the  deuteron}

\author{Carl E. Carlson}
\affiliation{Physics Department, College of William and Mary,
Williamsburg, VA 23187, USA}

\author{Marc Vanderhaeghen}
\affiliation{Physics Department, College of William and Mary,
Williamsburg, VA 23187, USA}

\affiliation{Institut f\"ur Kernphysik, Johannes Gutenberg-Universit\"at, D-55099 Mainz, Germany}

\date{\today}

\begin{abstract}
Using the recent empirical information on the deuteron electromagnetic 
form factors we map out the transverse charge density in the deuteron    
as viewed from a light front moving towards the deuteron.
The charge densities for a transversely polarized deuteron 
are characterized by monopole, dipole and quadrupole patterns.  
\end{abstract}

\pacs{13.40.Gp, 21.10.Ft, 21.45.Bc}

\maketitle
\thispagestyle{empty}

Electromagnetic form factors (FFs) of the deuteron have received a lot of 
attention in recent years; 
for recent reviews see {\it e.g.}  
Refs.~\cite{Garcon:2001sz,Gilman:2001yh}.
Based on the large amount of precise deuteron form factor data,  
it is therefore of interest to exhibit the spatial information on the 
quark charge distributions in the deuteron. 
In this letter, we develop the general formalism to extract 
charge densities for a spin-1 particle and apply it to the case of the 
deuteron. 

In the following we consider the electromagnetic deuteron elastic 
FFs when viewed from a light front moving towards the deuteron. 
Equivalently, this corresponds with 
a frame where the deuterons have a 
large momentum-component along the $z$-axis chosen along the direction of 
$P = (p + p^\prime)/2$, where $p$ ($p^\prime$) are the intial (final) 
deuteron four-momenta. We indicate the deuteron 
light-front + component by $P^+$ 
(defining $a^\pm \equiv a^0 \pm a^3$). 
We furthermore choose a light-front frame where the virtual photon 
four-momentum $q$ has $q^+ = 0$, 
and has a transverse component (lying in the $xy$-plane)
indicated by the transverse vector $\vec q_\perp$, satsifying 
$q^2 = - {\vec q_\perp}^{\, 2} \equiv - Q^2$. 
In such a light-front frame, the virtual photon only couples to forward moving 
partons and the + component of the 
electromagnetic current $J^+$ 
has the interpretation of the quark charge density operator. It is  
given by  $J^+(0) = +2/3 \, \bar u(0) \gamma^+ u(0) - 1/3 \, 
\bar d(0) \gamma^+(0) d(0)$, considering only $u$ and $d$ quarks. 
Each term in the expression is a positive operator 
since $\bar q \gamma^+ q \propto | \gamma^+ q |^2$. 

In the following, we will use empirical
information on deuteron elastic FFs to study the 
deuteron quark charge densities in the transverse plane.   
It is customary to denote the three 
deuteron elastic e.m.~FFs by $G_C$, $G_M$, and $G_Q$,  
corresponding to the Coulomb monopole ($G_C$), magnetic dipole ($G_M$), 
and Coulomb quadrupole ($G_Q$) FFs respectively.   Similar relations between
nucleon densities and FFs can be found 
in~\cite{Pasquini:2007xz,Miller:2007uy,Carlson:2007xd,Miller:2007kt}.

We start by expressing the matrix elements of the $J^+(0)$ operator 
between deuteron states as,
\begin{eqnarray}
\langle P^+, \frac{\vec q_\perp}{2}, \lambda^\prime | J^+(0) | 
P^+, - \frac{\vec q_\perp}{2}, \lambda  \rangle 
&=& (2 P^+)  e^{i  (\lambda - \lambda^\prime) \phi_q} 
\nonumber \\
&\times& \, G^+_{\lambda^\prime \, \lambda} (Q^2),
\label{eq:dens1}
\end{eqnarray}
where $\lambda$~=~$\pm 1,0$ ($\lambda^\prime$~=$~\pm 1, 0$) 
denotes the initial (final) deuteron light-front helicity, 
and where 
$\vec q_\perp = Q ( \cos \phi_q \,\hat e_x + \sin \phi_q \,\hat e_y )$. 
Furthermore in Eq.~(\ref{eq:dens1}), the helicity form factors 
$G^+_{\lambda^\prime \, \lambda}$ are real (due to time reversal 
invariance) and depend only on $Q^2$.

We can then define transverse charge densities for deuteron in helicity 
states of $\lambda = \pm 1$ or $\lambda = 0$ as,
\begin{eqnarray}
\rho^d_{\lambda} (b) 
\!\!&\equiv& \!\!\int \frac{d^2 \vec q_\perp}{(2 \pi)^2} 
e^{- i \, \vec q_\perp \cdot \vec b}  \frac{1}{2 P^+} 
\langle P^+, \frac{\vec q_\perp}{2}, \lambda 
| J^+ | P^+, \frac{- \vec q_\perp}{2}, \lambda  \rangle  
\nonumber \\
\!\!&=&\!\! \int_0^\infty \frac{d Q}{2 \pi} Q \, 
J_0(b \, Q) \, G^+_{\lambda \lambda}(Q^2),
\label{eq:dens2}
\end{eqnarray}
where $\vec b \equiv b \, 
(\cos \phi_b \, \hat e_x + \sin \phi_b \, \hat e_y)$  
denotes the position in the $xy$ plane 
from the transverse {\it c.m.} of the deuteron. 
The two independent helicity conserving FFs 
$G^+_{1 \, 1}$ and $G^+_{0 \, 0}$ can be expressed in terms of 
$G_{C, M, Q}$ as, 
\begin{eqnarray}
G^+_{1 \, 1} &=& \frac{1}{1 + \eta} 
\left \{G_C + \eta \, G_M + \frac{\eta}{3} \, G_Q \right \} , 
\\
G^+_{0 \, 0} &=& \frac{1}{1 + \eta} 
\left \{(1 - \eta) \, G_C + 2 \eta \, G_M - 
\frac{2 \eta}{3} (1 + 2 \eta) G_Q \right \} , \nonumber 
\end{eqnarray}
with $\eta \equiv Q^2 / (4 M_d^2)$, 
and $M_d$ is the deuteron mass.    

The definitions and normalizations of $G_C$, $G_M$, and $G_Q$ are the customary ones obtained from the matrix elements of the electromagnetic current~\cite{Arnold:1979cg},
\begin{eqnarray}
&&\hspace{-0.4cm}
\left\langle p_2,\lambda_2 \right|  J^\mu  \left| p_1, \lambda_1 \right\rangle 
= -  (\varepsilon_2^* \cdot \varepsilon_1) 2P^\mu G_1(Q^2)
			\\
&&\hspace{-0.4cm}
- \left( \varepsilon_1^\mu  \varepsilon_2^* \cdot q
- {\varepsilon_2^\mu}^*  \varepsilon_1 \cdot q \right) G_M(Q^2)
+ q \cdot \varepsilon_1 \, q\cdot \varepsilon_2^* \,
		\frac{P^\mu}{M_d^2} G_3(Q^2), \nonumber 
\end{eqnarray}
with $\varepsilon_{1,2}$ the polarization vectors of the initial and final deuteron.   The charge and quadrupole FFs follow from,
\begin{eqnarray}
G_C &=& G_1 + \frac{2}{3} \eta \, G_Q, 	
\nonumber \\ 	
G_Q &=& G_1 + \left(1+ \eta\right) G_3 - G_M ,
\end{eqnarray}
with normalizations $G_C(0)=1$, $G_M(0) = \mu_d$ [magnetic moment 
in units $e/(2M_d)$], and $G_Q(0) = Q_d$ (quadrupole moment in 
units $e/M_d^{2}$).

For numerical evaluation, we use the parameterization of the deuteron form factor data given as fit II by Abbott {\it et al.}~\cite{Abbott:2000ak}.  This parameterization is based on forms suggested in~\cite{Kobushkin:1994ed}, which read
\begin{align}
& G_C = \frac{ G^2( \frac{Q^2}{4}) }{ 2\eta +1 } 
	\left[ \big( 1-\frac{2}{3} \eta \big) g_0
		+ \frac{8}{3} \sqrt{2\eta} \, g_1 + \frac{2}{3} ( 2\eta -1 ) g_2     \right] \!  ,
			\nonumber \\
& G_M = \frac{ G^2( \frac{Q^2}{4}) }{ 2\eta +1 } 
	\left[ 2 g_0
		+ \frac{ 2 ( 2\eta -1 ) }{\sqrt{2\eta}}  g_1 - 2 g_2           \right]   ,
			\nonumber \\
& G_Q = \frac{ G^2( \frac{Q^2}{4}) }{ 2\eta +1 } 
	\left[ - g_0
		+ \sqrt{ \frac{2}{\eta} }  g_1 - \frac{ \eta+1 }{ \eta } g_2   \right]   ,
\end{align}
where the dipole form factor $G(Q^2) = (1 + Q^2 / \delta^2 )^{-2}$ has a (non-standard) mass parameter $\delta = 898.52$ MeV.   The reduced amplitudes are 
\begin{eqnarray}
g_0 &=& \sum_{i=1}^4  \frac{ a_i }{ \alpha_i^2 + Q^2 }  ,  \quad \quad
g_1 = Q \sum_{i=1}^4  \frac{ b_i }{ \beta_i^2 + Q^2 }  ,   \nonumber \\
g_2 &=& Q^2 \sum_{i=1}^4  \frac{ c_i }{ \gamma_i^2 + Q^2 }  ,
\end{eqnarray}
and the values of the parameters obtained by the authors of~\cite{Abbott:2000ak} are given in Table~\ref{table:abbottfit}.

\begin{table}[h]
\begin{center}

\begin{ruledtabular}
\begin{tabular}{|c|c|c|c|c|c|}
$a_i$	& $b_i$	& $c_i$	& $\alpha_i^2$	& $\beta_i^2$     & $\gamma_i^2$       \\
(fm$^{-2}$)  & (fm$^{-1}$)	& & (fm$^{-2}$)	& (fm$^{-2}$) & (fm$^{-2}$)    \\
\hline
1.57057		& 0.07043	& $-0.16577$	& 1.52501	& 43.67795   & 1.87055  \\
12.23792		& 0.14443	& 0.27557	& 8.75139	& 30.05435   & 14.95683 \\
$-42.04576$	& $-0.27343$	& $-0.05382$	& 15.97777	& 16.43075   & 28.04312  \\
$27.92014$	& $0.05856$	& $-0.05598$	& $23.20415$	& $2.80716$ & $41.12940$  \\
\end{tabular}
\end{ruledtabular}
\end{center}
\caption{Parameters for deuteron form factor fit II of Ref.~\cite{Abbott:2000ak}. }

\label{table:abbottfit}
\end{table}%

\begin{figure}[h]
\vspace{-0.25cm}
\begin{center}
\includegraphics[width =6.5cm]{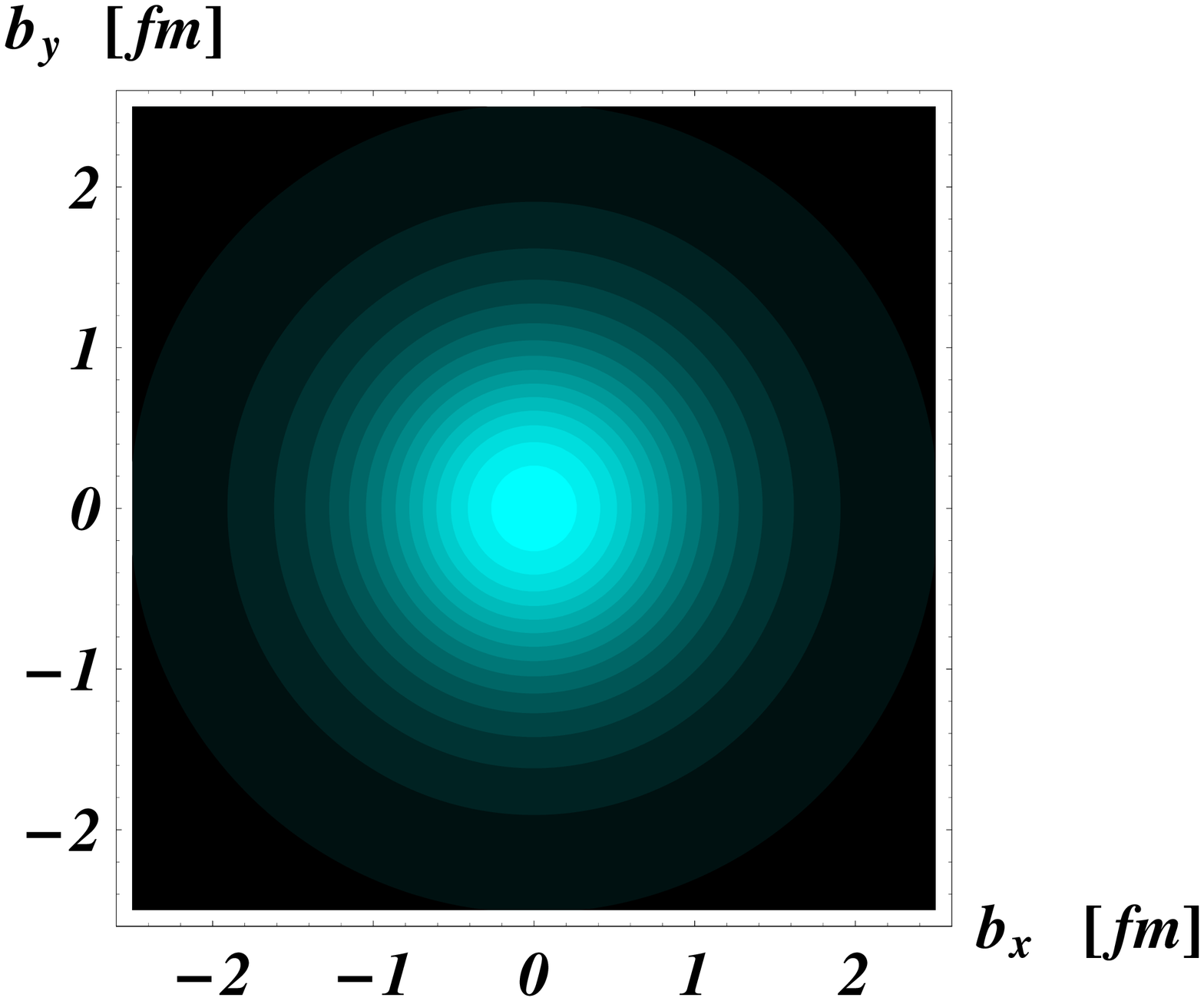}
\includegraphics[width =6.5cm]{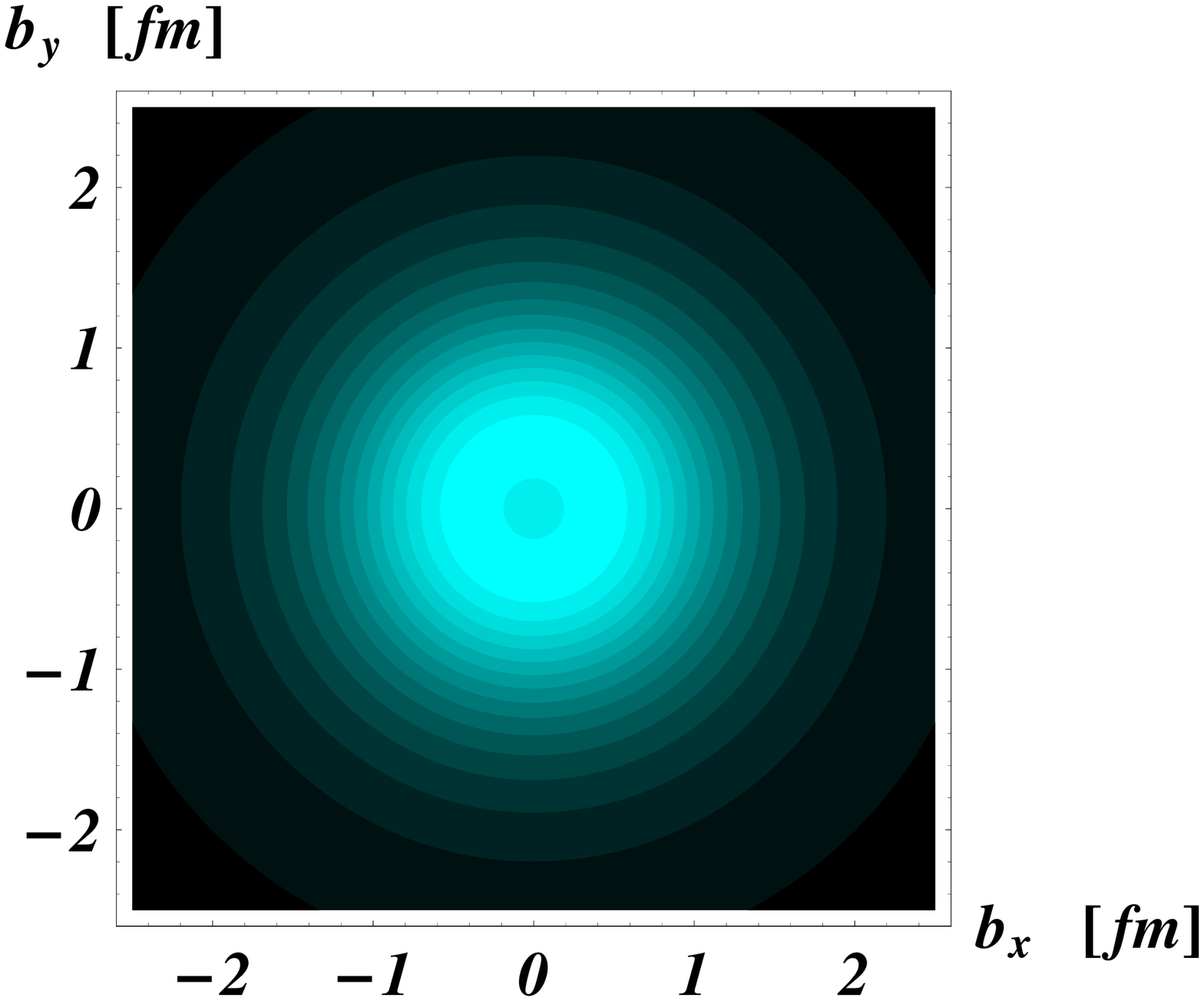}
\includegraphics[width =6.5cm]{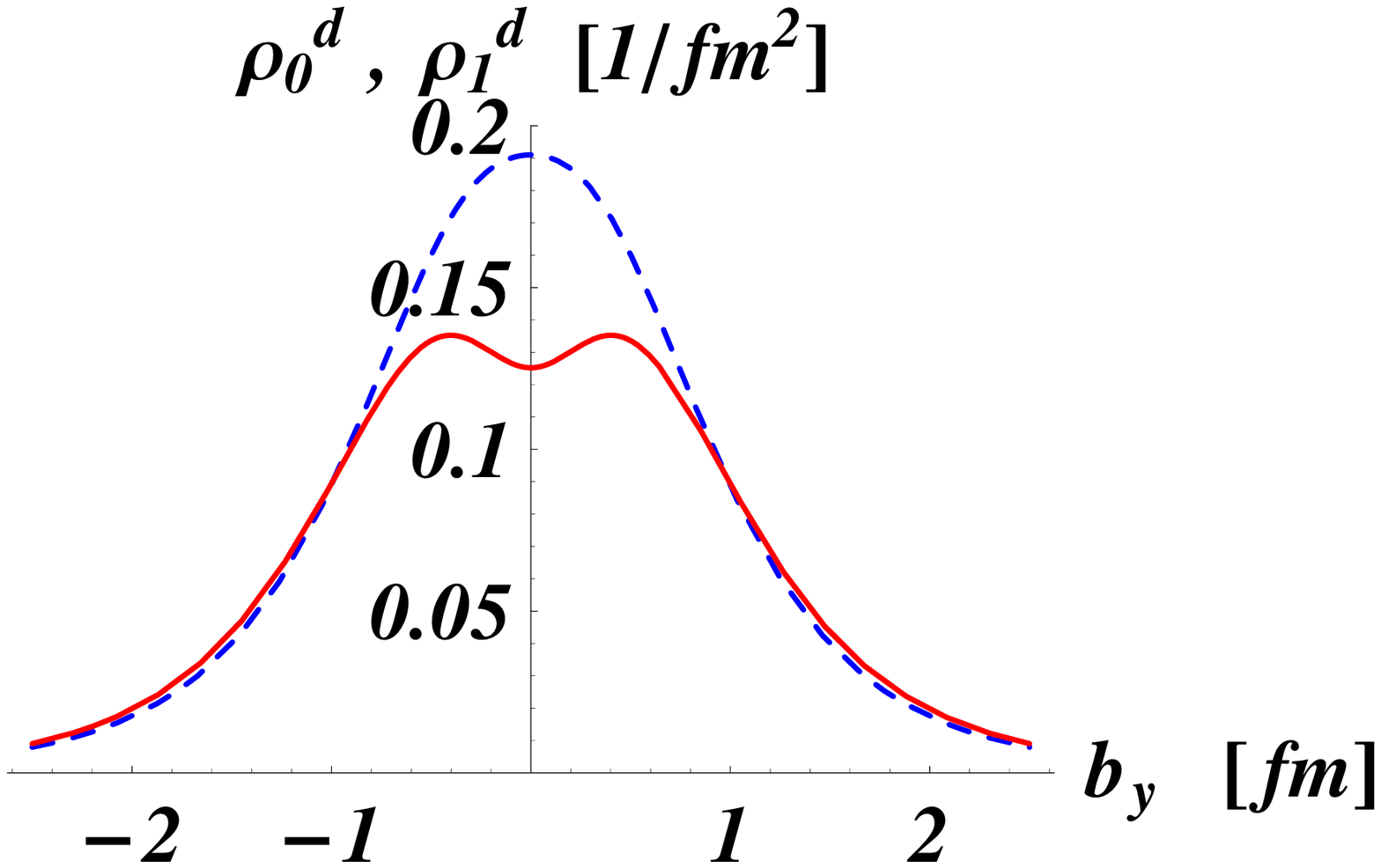}
\end{center}
\caption{Quark transverse charge densities in the {\it deuteron}. 
Upper panel : $\rho^d_1$. Middle panel : $\rho^d_0$. 
The lightest (darkest) regions correspond with
largest (smallest) densities. 
Lower panel : the density along the $y$-axis 
for  $\rho^d_1$ (dashed curve) and $\rho^d_0$ (solid curve). 
For the deuteron FFs, we use the empirical parameterization II of 
Ref.~\cite{Abbott:2000ak}, based on Ref.~\cite{Kobushkin:1994ed}. }
\label{fig:deuteronlongpol}
\end{figure}

Figure~\ref{fig:deuteronlongpol} shows the transverse charge densities for definite helicity (longitudinally polarized) deuterons.  Recall that the transverse charge density is the charge density projected onto a plane perpendicular to the line of sight, which also defines the longitudinal direction.  The charge density for the $\lambda = 1$ state, in the upper panel, is smooth and peaks in the center.  The $\lambda = 0$ state, in the middle panel, features a dip in the center, which can also be seen in the lower panel, where the density along the $y$-axis is plotted for both states.  

This dip---found directly from the data---reflects the low charge density or ``hole'' in the center of the deuteron wave function.  For the helicity-0 state, calculated S- and D- state wave functions produce toroidal equidensity surfaces in the central regions, with the axis along the $z$ direction~\cite{Forest:1996kp}.  Hence looking along the $z$-axis for this state, one can see the hole in the center, albeit the hole be partly filled in because one must look though the outer layers of the deuteron which are mostly S-state and spherically symmetric.  We emphasize that the dip seen in our figure is based only on observation (codified in the form factor fits) and a light-front interpretation of the data.  The result is consistent with but does not use calculations based on models of nucleon-nucleon interactions.  Further, the light front viewpoint gives a two-dimensional charge density that is relativistically correct, unlike charge densities obtained from Fourier transforming form factors in an equal time formalism.

We next consider the charge densities for a transversely polarized deuteron, 
denoting the transverse polarization direction by 
$\vec S_\perp = \cos \phi_S \,\hat e_x + \sin \phi_S \,\hat e_y$. 
The transverse charge densities can be defined as,
\begin{eqnarray}
\rho^d_{T \, s_\perp} (\vec b) 
&\equiv& \int \frac{d^2 \vec q_\perp}{(2 \pi)^2} \,
e^{- i \, \vec q_\perp \cdot \vec b} \, \frac{1}{2 P^+} 
\label{eq:dens4} \\
&& \hspace{1.25cm} \times   
\langle P^+, \frac{\vec q_\perp}{2}, s_\perp 
| J^+ | P^+, \frac{- \vec q_\perp}{2}, s_\perp  \rangle, 
\nonumber 
\end{eqnarray}
where $s_\perp$ is the deuteron  
spin projection along the direction of $\vec S_\perp$.
The transverse spin state can be expressed in terms of the light
front helicity spinor states as, 
\begin{eqnarray}
| s_\perp = \pm 1 \rangle &=& \frac{1}{2}
\Big(  | \lambda = + 1 \rangle 
\pm \sqrt{2} e^{i \phi_S } \, | \lambda = 0 \rangle
					\nonumber \\
&& \qquad + \ e^{2i \phi_S } \, | \lambda = - 1\rangle \Big)    ,
					\nonumber \\
| s_\perp = 0 \rangle &=& \frac{1}{\sqrt{2}}
\Big(  | \lambda = + 1 \rangle 
	- \ e^{2i \phi_S } \, | \lambda = - 1\rangle \Big)    .
\end{eqnarray} 

By working out the Fourier transform in Eq.~(\ref{eq:dens4}) for the 
two cases where $s_\perp = +1$ and $s_\perp = 0$, one obtains,
\begin{eqnarray}
\label{eq:dens5} 
\rho^d_{T \, 1} (\vec b) & = & \int_0^\infty \frac{d Q}{2 \pi} \, 
Q\, \left\{ 
J_0(b \, Q) \, \frac{1}{2} \left( G^+_{1 \, 1} + G^+_{0 \, 0} \right)
\right. \nonumber \\  
&&\hspace{0.5cm} 
+ \sin(\phi_b - \phi_S) \, J_1(b \, Q) \sqrt{2} \,  G^+_{0 \, 1}  \\ \nonumber
&&\left. \hspace{0.5cm}- \cos 2 (\phi_b - \phi_S) \, J_2(b \, Q) 
\frac{1}{2} \, G^+_{-1 \, +1}  
\right\} , \\ 
\rho^d_{T \, 0} (\vec b) & = & \int_0^\infty \frac{d Q}{2 \pi} \, 
Q\, \left\{ J_0(b \, Q) \, G^+_{1 \, 1} \right. \nonumber \\  
&&\left. \hspace{0.5cm}+ \cos 2 (\phi_b - \phi_S) \, J_2(b \, Q) 
\, G^+_{-1 \, +1} \right\} . 
\label{eq:dens6}  
\end{eqnarray}
One notices from Eqs.~(\ref{eq:dens5},\ref{eq:dens6}) 
that the transverse charge density $\rho^d_{T \, 1}$ 
contains monopole, dipole and quadrupole field patterns, whereas 
$\rho^d_{T \, 0}$ 
only contains monopole and quadrupole field patterns. 
The deuteron helicity FF with one unit of helicity flip, 
which governs the dipole field pattern in
$\rho^d_{T \, 1}$ , can be expressed in terms of $G_{C, M, Q}$ as, 
\begin{eqnarray}
G^+_{0 \, 1} &=& - \frac{\sqrt{2 \eta}}{1 + \eta} 
\left \{G_C - \frac{1}{2} (1 - \eta) \, G_M + \frac{\eta}{3} \,G_Q \right \} , 
\label{eq:dens7}
\end{eqnarray}
whereas the deuteron helicity FF with two units of helicity flip, 
corresponding with the quadrupole field patterns, can be expressed as, 
\begin{eqnarray}
G^+_{-1 \, +1} &=& \frac{\eta}{1 + \eta} 
\left \{G_C - G_M - (1 + \frac{2 \eta}{3}) \,G_Q \right \} . 
\label{eq:dens8}
\end{eqnarray}
Note that the four deuteron helicity FFs introduced here are not 
independent. The angular condition~\cite{Carlson:2003je} relating them reads,
\begin{eqnarray}
(2 \eta + 1) G^+_{1 \, 1} + 2 \sqrt{2 \eta} \, G^+_{0 \, 1} + 
G^+_{-1 \, +1} - G^+_{0 \, 0} = 0.
\end{eqnarray}

It is instructive to evaluate the electric dipole moment (EDM)  
corresponding with the transverse charge densities $\rho^d_{T \, s_\perp}$, 
which is defined as, 
\begin{eqnarray}
\vec d^d_{s_\perp} \equiv e \int d^2 \vec b \, \vec b \, 
\rho^d_{T \,  s_\perp}(\vec b). 
\end{eqnarray}
Eq.~(\ref{eq:dens6}) gives $\vec d^d_{0} = 0$, 
whereas Eq.~(\ref{eq:dens5}) yields, 
\begin{eqnarray}
\vec d^d_{1} = - \left( \vec S_\perp \times \hat e_z \right) 
\, \left\{ G_M(0) - 2 \right\} \, \left( \frac{e}{2 M_d} \right). 
\label{eq:edm}
\end{eqnarray}
Expressing the spin-1 magnetic moment in terms of the $g$-factor,   
{\it i.e.} $G_M(0) = g$, 
one sees that the induced EDM  
$\vec d^d_1$ is proportional to $g - 2$. The same result was found   
for the case of a spin-1/2 particle~\cite{Carlson:2007xd}.
One thus observes that for a particle 
without internal structure (corresponding with $g = 2$~\cite{Ferrara:1992yc}), 
there is no induced EDM.  

One can also evaluate the electric quadrupole moment corresponding 
with the transverse charge densities $\rho^d_{T \, s_\perp}$.  
Choosing $\vec S_\perp = \hat e_x$, the electric quadrupole 
moment can be defined as,  
\begin{eqnarray}
Q^d_{s_\perp} \equiv e \int d^2 \vec b \, (b_x^2 - b_y^2) \, 
\rho^d_{T \,  s_\perp}(\vec b). 
\end{eqnarray}
From Eqs.~(\ref{eq:dens5},\ref{eq:dens6}) 
one obtains, 
\begin{eqnarray}
Q^d_{1} &=& - \frac{1}{2} Q^d_{0} \nonumber \\
&=& \frac{1}{2} 
 \left\{ \left[ G_M(0) - 2 \right] + 
\left[ G_Q(0) + 1 \right] \right\} \left( \frac{e}{M_d^2} \right) . 
\label{eq:quadrup}
\end{eqnarray}
\indent
We may note that for a spin-1 particle without internal structure, such as the 
$W$ and $Z$ gauge bosons in the standard electroweak theory, it is  
required that at tree level $G_M(0) = 2$ and $G_Q(0) = -1$, in order to 
satisfy the Gerasimov-Drell-Hearn sum rule to lowest order in perturbation 
theory~\cite{Kim:1973ee,Brodsky:1992px}.   
For the gauge bosons, any deviations from these values would indicate new 
(beyond standard model) physics. Current data from $p \bar p$ collisions 
(up to $\sqrt{s} = 1.96$~TeV) are compatible with zero 
anomalous $WW \gamma$ and $Z Z \gamma$ 
couplings~\cite{Abazov:2008vja}.

It is thus interesting to observe from Eq.~(\ref{eq:quadrup}) 
that $Q^d_{s_\perp}$ is only sensitive to the anomalous parts 
of the spin-1 magnetic dipole and electric quadrupole moments, 
and vanishes for a particle without internal structure. 
The deuteron has a magnetic dipole moment $G_M(0) = 1.71$~\cite{Mohr:2008},  
close to the natural value for a spin-1 particle. However,  
in contrast to the $W$ and $Z$ gauge bosons, the deuteron has a large 
anomalous quadrupole moment. 
Its measured value is $G_Q(0) = 25.84 \, (3)$~\cite{Ericson:1982ei}, 
highlighting the prominent role of the pion exchange potential. 
Eq.~(\ref{eq:quadrup}) thus shows that for a deuteron polarized along the 
$x$-axis, its transverse charge densities show  
large quadrupole moments ($Q^d_1 > 0$ and $Q^d_0 < 0$).   

\begin{figure}[t]
\vspace{-0.25cm}
\begin{center}
\includegraphics[width =6.5cm]{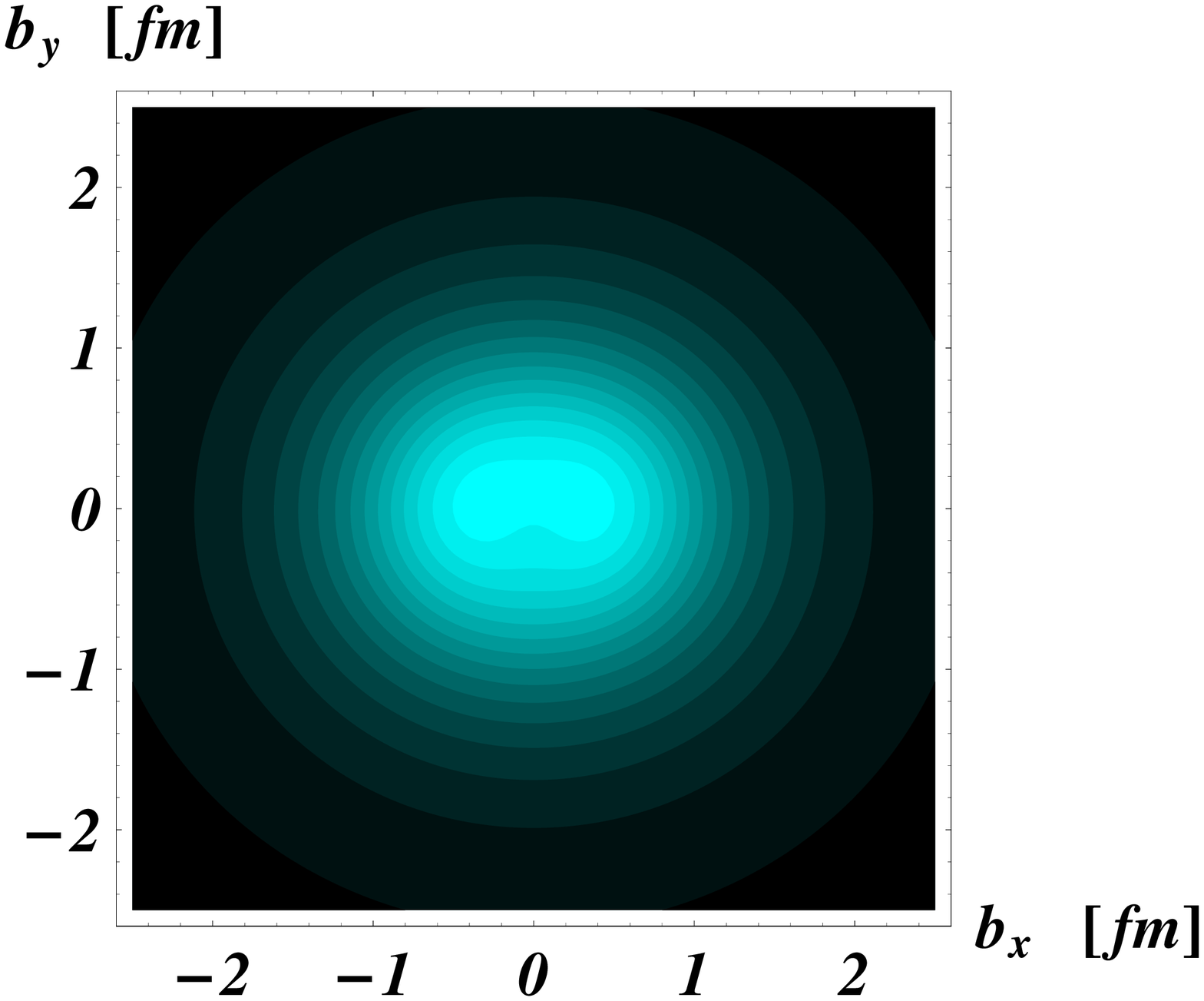}
\includegraphics[width =6.5cm]{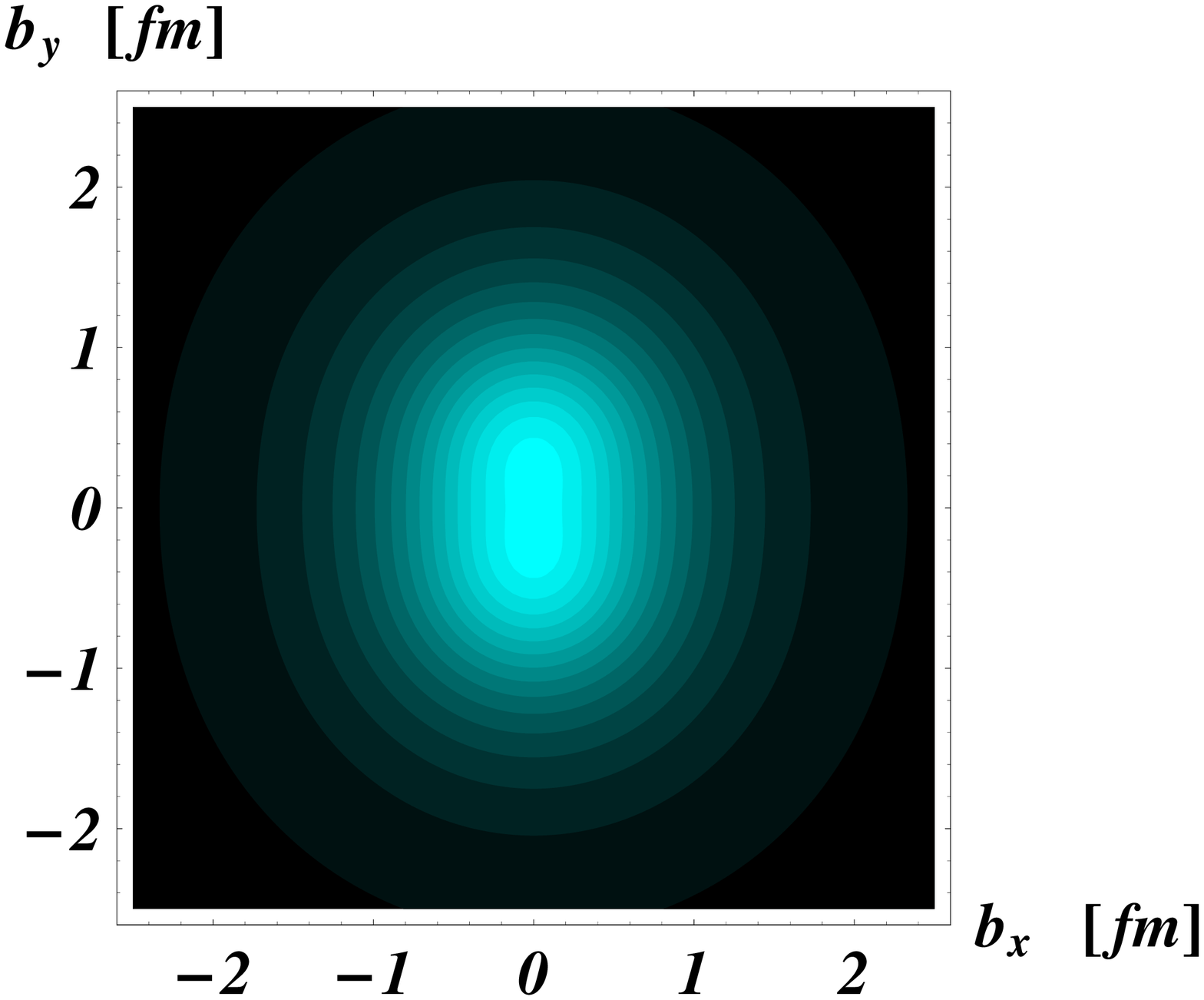}
\includegraphics[width =6.5cm]{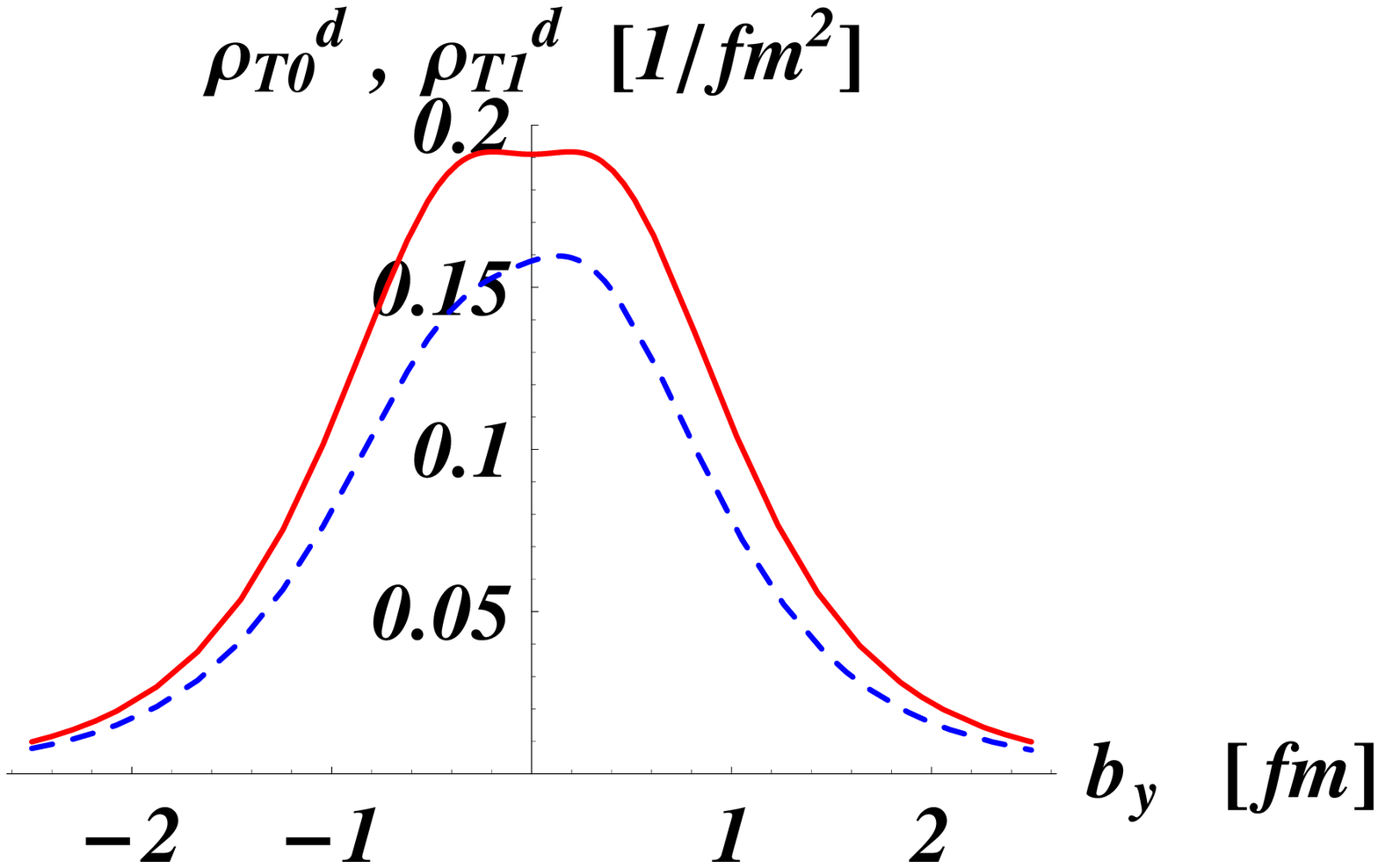}
\end{center}
\caption{Quark transverse charge densities for a {\it deuteron} 
which is polarized along the positive $x$-axis. 
Upper panel : $\rho^d_{T \, 1}$. Middle panel : $\rho^d_{T \, 0}$. 
The light (dark) regions correspond with
largest (smallest) densities. 
The lower panel compares the density along the $y$-axis 
for  $\rho^d_{T \, 1}$ (dashed curve) and $\rho^d_{T \, 0}$ (solid curve). 
For the deuteron e.m. FFs, we use the empirical parameterization II of 
Ref.~\cite{Abbott:2000ak}, based on Ref.~\cite{Kobushkin:1994ed}. }
\label{fig:deuterontrans}
\end{figure}

Pictorial results for the transverse charge density with transverse deuteron
polarization are shown in Fig.~\ref{fig:deuterontrans}.   The upper panel
shows the charge distribution for the state with projection-1 in the
$x$-direction. One sees the large effects of the quadrupole term together 
with a small overall shift of the charge distribution 
along the $y$-axis.  
The latter is consonant with the observation~\cite{Einstein:1908,Krotkov99} 
that an object with a magnetic dipole moment when stationary exhibits an 
electric dipole moment when moving, or when seen by a moving observer, 
proportional to the vector product $\vec v \ \times$ (magnetic moment).  
The middle panel shows charge density for a state with projection-0 in 
the $x$-direction.  The quadrupole term stretches the charge along 
the $y$-axis but does not cause any shift of the charge center.  
The two cases are compared in the lower panel, which plots the density 
along the $y$-axis.

In summary, we used recent empirical information on the deuteron 
e.m.~FFs to map out the transverse charge densities in longitudinally 
and transversely polarized deuterons.    Notably, one sees a dip in 
the center of the charge distribution for helicity-zero deuterons.  
This is in harmony with nuclear force model calculations, 
which for the zero helicity case predict toroidal equidensity surfaces 
for higher densities, with axis along the quantization direction.  
Transversely polarized deuterons show dipole and quadrupole structure 
in the charge distributions. Their electric dipole and quadrupole moments 
only depend on the spin-1 particle's anomalous magnetic dipole moment and its 
anomalous electric quadrupole moment, arising from its internal structure.

\begin{acknowledgments} 
The work of C.~E.~C. is supported by the NSF under grant PHY-0555600. 
The work of M.~V. is supported in part by DOE grant
DE-FG02-04ER41302.
\end{acknowledgments}

\end{document}